\pdfoutput=1
\documentclass[english]{article}
\usepackage[T1]{fontenc}
\usepackage[latin9]{inputenc}
\usepackage{geometry}
\usepackage{textcomp}
\usepackage{amsmath}
\usepackage{amssymb}
\usepackage{graphicx}
\usepackage[superscript,biblabel]{cite}
\usepackage{setspace}
\usepackage{float}
\usepackage{upgreek}
\makeatletter
\newcommand{\lyxaddress}[1]{
\par {\raggedright #1
\vspace{1.4em}
\noindent\par}
}

\makeatother

\usepackage{babel}
\begin{document}

\title{Classical approach to collision complexes in ultracold chemical reactions}

\author{Micheline B. Soley, Eric J. Heller}
\maketitle

\lyxaddress{Department of Chemistry and Chemical Biology, Harvard University,
Cambridge, Massachusetts, 02138, USA}
\begin{abstract}
Inspired by Wannier's threshold law,\cite{Wannier.1953.817} we recognize that collision complex decay meets the requirements of quantum-classical correspondence in sufficiently exothermic ultracold reactions.  We make use of this correspondence to elucidate the classical foundations of ultracold reactions and to help bring calculations currently beyond the capabilities of quantum mechanics within reach.  A classical method with a simplified model of many-body interactions is provided for determination of the collision complex lifetime and demonstrated for a reduced-dimensional system, as preliminary to the calculation of collision complex lifetimes in the full-dimensional system.

\end{abstract}

\section{Introduction}

\begin{figure}[H]
\begin{centering}
\includegraphics[width=0.5\textwidth]{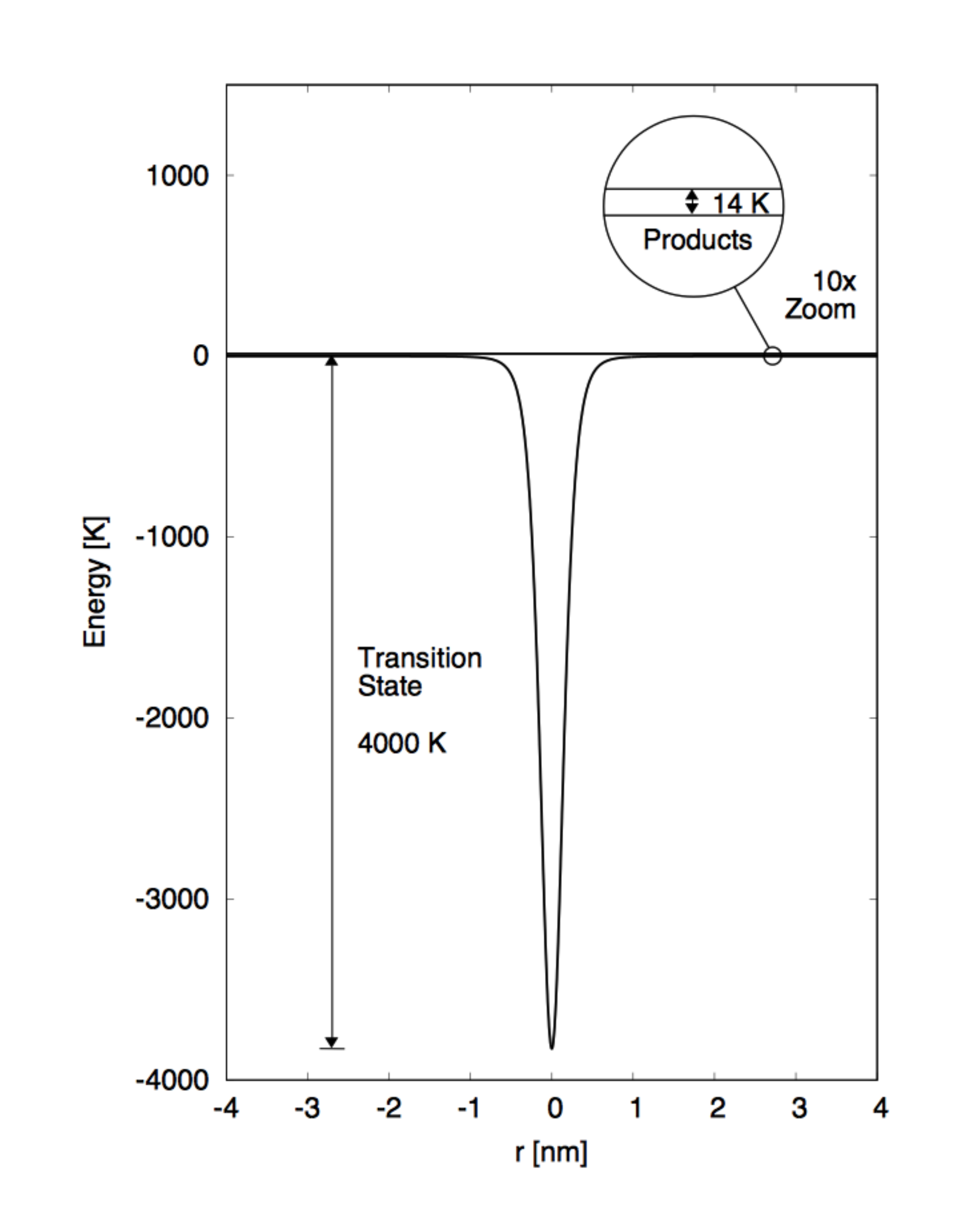}
\par\end{centering}
\caption{Extreme energy differences in the ultracold $\text{KRb}$ dimer reaction collision complex raise the question of whether classical or quantum effects dominate in the decay process. Magnification of the circle diameter and energy lines are shown to scale with x10 magnification.
\label{fig:Energetics}}
\end{figure}

With temperatures below a milliKelvin, ultracold chemical reactions are under investigation as the means to form product molecules with unprecedented specificity.\cite{Ospelkaus.2010}  This provides opportunities to investigate the mechanisms behind chemical reactions\cite{Ospelkaus.2010} and to "tune" reactions to produce desired products with coherent control.\cite{Krems.2005,Krems.2008}  Formation of products from a collision complex in the ultracold potassium-rubidium $\text{KRb}$ dimer reaction is the subject of ongoing experimental and theoretical work:\cite{Ni.2008,Hutson.2010,Ni.2010,Ospelkaus.2010,deMiranda.2011.502,DeMarco.2018.1808.00028v1}
\begin{equation}
2\text{KRb}\rightarrow\left[\text{K}_{2}\text{Rb}_{2}\right]^{\star}\rightarrow\text{K}_{2}+\text{Rb}_{2}.\label{eq:Reaction}
\end{equation}
One of the first ultracold chemical reactions to be carried out experimentally,\cite{Ni.2008,Hutson.2010,Ni.2010,Ospelkaus.2010} the ultracold $\text{KRb}$ dimer reaction is interesting to study due to the extreme energy differences involved.   In the reaction, two ultracold $\text{KRb}$ dimers meet at approximately $300\text{ nK}$ to form a transition-state collision complex $\left[\text{K}_{2}\text{Rb}_{2}\right]^{\star}$. The collision complex, which we refer to as a "cauldron," is energetically favorable by $4000\text{ K}$.\cite{Byrd.2010}  The atoms in the cauldron experience four-body interactions and are expected to move chaotically. Ultimately, the collision complex breaks apart to form cold potassium $\text{K}_{2}$ and rubidium $\text{Rb}_{2}$ dimers. The equivalent of $14\text{ K}$\footnote{The ultracold $\text{KRb}$ reaction occurs without coupling to a bath, such that the product ensemble is expected to be near microcanonical and the products are not suspected to be distributed thermally.} is expected to be released.\cite{Ospelkaus.2010,Ni.2008,Amiot.1990,Falke.2006} The reaction therefore has an ultracold approach followed by a hot "cauldron" and a cold departure.  This is key because the lifetime of the collision complex remains unknown.   The hot cauldron and the cold departure depicted in Fig.~\ref{fig:Energetics} raises the question -- will classical or quantum effects dominate the break-up of the collision complex?

This paper involves a simplified two-dimensional model of the ultracold $\text{KRb}$ dimer reaction in which product dimers are considered to be structureless point particles.  This two-dimensional system retains the essence of much of the full four-atom case: the hot cauldron of the transition state and the cold departure of the products.  While the $300\text{ nK}$ temperature of the reactants is ultracold, having a temperature below $1\text{ mK}$, the $14\text{ K}$ released in formation of the products is well above the ultracold regime. Surprisingly, whereas the ultracold temperature of the reaction would typically indicate a need for quantum mechanics, semiclassical arguments suggest that given the exothermicity of the reaction, for the two-dimensional case, the classical rate will be nearly exact. Furthermore, this correspondence is not specific to collision complex decay in the ultracold $\text{KRb}$ dimer reaction, but holds more generally for sufficiently exothermic bimolecular ultracold reactions.  In these cases, classical mechanics can be substituted where quantum mechanics becomes intractable. By determining the collision complex decay rate for the two-dimensional system classically, we solve a key problem encountered in calculation of the collision complex decay rate for the ultracold $\text{KRb}$ dimer reaction.

Even in the two-dimensional case, it is difficult to simulate collision complex decay in the ultracold $\text{KRb}$ dimer reaction quantum mechanically.  The collision complex supports a large number of states, which requires a large basis set expansion.  The collision complex also breaks apart into cold products with long de Broglie wavelengths, which require grid techniques that take into account a wide expanse of position space.  The vast, sudden change in scale between representation of the hot collision complex cauldron and the cold products is a well-known danger signal for quantum mechanical calculations.  These concerns, as well as the exponential growth of the Hilbert space with system size, impede rigorous quantum mechanics from being used to simulate ultracold reactions of four heavy atoms. In contrast, classical methods are available for simulations of systems of many heavy atoms, given de Broglie wavelengths in the classical regime.  

Studies of the ultracold $\text{KRb}$ dimer reaction have focused on prediction and analysis of the experimental reactant loss rate.  These studies have employed the quantum threshold model,\cite{Quemener.2010.022702,Quemener.2010.060701R,Quemener.2011.062703,Gonzalez-Martinez.2014.052716} the quantum Langevin model,\cite{Gao.2010.263203} the statistical adiabatic channel model,\cite{Buchachenko.2012.114305} quantum defect theory,\cite{Julienne.2011.19114,Wang.2012.062704} multichannel quantum defect theory\cite{Gao.2010.263203,Idziaszek.2010.113202,Idziaszek.2010.020703R}, and other time-independent quantum mechanical formalisms.\cite{Quemener.2011.012705,Wang.2015.035015}. The quantum Rice-Ramsperger-Kassel-Marcus (RRKM) decay rate has also been calculated for the collision complex as a step towards calculation of the scattering cross-sections using time-independent multi-channel quantum defect theory and random matrix theory.\cite{Mayle.2013.012709} It is fair to say that studies of the collision complex decay are still at the qualitative level.  To produce quantitative results, new methods are needed that can circumvent the difficulties posed by quantum mechanics.

We introduce the concept of quantum-classical correspondence to collision complex decay in sufficiently exothermic ultracold reactions and present a classical method based on Wannier's threshold law.\cite{Wannier.1953.817}   In his seminal paper on electrical discharge from gases, Wannier was concerned with the double escape of electrons from an ion.  Recognizing the de Broglie wavelengths at long-range were in the classical regime, Wannier calculated the energy-dependence of the collision complex decay rate classically.  Integration over the phase-space configurations leading to product formation yielded the threshold law $E^{1.127}$.  This threshold law was subsequently confirmed semiclassically,\cite{Peterkop.1971.513} quantum mechanically,\cite{Rau.1971.207} and experimentally.\cite{Cvejanovic.1974.1841} The two-dimensional model of collision complex decay in the ultracold $\text{KRb}$ dimer reaction, which we are presenting here, is a similarly special system.

\begin{figure}[H]
\begin{centering}
\includegraphics[width=0.9\textwidth]{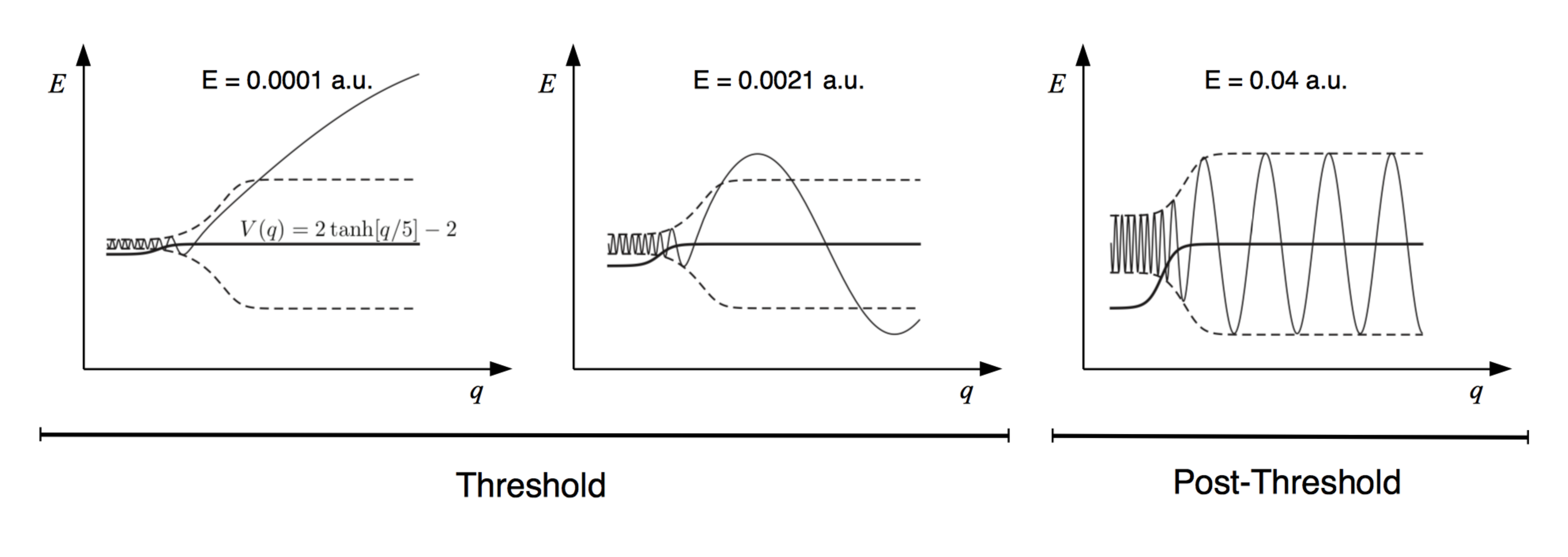}
\par\end{centering}
\caption{(a,b) Representative quantum eigenstates (thin solid line) differ from WKB semiclassical amplitudes (thin dashed line) in the threshold region but (c) agree in the post-threshold regime, as shown for the potential energy surface $V(q)=2\tanh[q/5]-2$ (thick solid line).\cite{Heller}  WKB amplitudes give classical probabilities, such that classical results are also nearly equal to quantum results in the post-threshold regime. Energies $E$ are given in atomic units and the potential is rescaled between frames to highlight details.
\label{fig:PostThreshold}}
\end{figure}

First, in both Wannier's system and the $\text{KRb}$ system presented here, only one classical pathway leads to each final state along the reaction coordinate.  When there is only one path, the sum of the square root of the classical probability density for each contributing classical path in the semiclassical amplitude reduces to one term.  There is then no semiclassical interference and the semiclassical and classical probabilities agree.   Second, there is no quantum reflection, a quantum mechanical effect in which particles are reflected from attractive potentials at low energies.  In Wannier's system, there is no quantum reflection threshold at any energy,\cite{Wannier.1953.817,Brenig.1992.397,Carraro.1995} as the Coulomb potential is a special case for which the Wentzel-Kramers-Brillouin (WKB) criterion is always satisfied (i.e. the particle wavelength is small in comparison to the rate of change of the potential energy surface and, equivalently, the wavelength changes little in the length of a wavelength).\cite{Wentzel.1926.518,Kramers.1926.828,Brillouin.1926.24,Brillouin.1926.353,Cote.1996,Mody.2001} In the $\text{KRb}$ system, enough energy is released that the system is in what we term the "post-threshold" regime above the quantum reflection threshold for the potential,\cite{Forrey.1999.2657,Mody.2001,Carraro.1995}  which implies that at the exothermicity of the reaction, the WKB criterion is satisfied at all positions along the reaction coordinate.  The semiclassical approximation then holds such that the semiclassical wavefunctions closely approximate the quantum wavefunctions, as illustrated in Fig.~\ref{fig:PostThreshold}(c) for a ramp potential.  In tandem, these two conditions make it such that the classical probabilities are nearly equal to the quantum probabilities.

The quantum-classical correspondence allows us to make a classical interpretation of how collision complex decay occurs in the two-dimensional model.  From a classical perspective, product formation from the collision complex is a rare event.  Although the decay process is barrierless and exothermic, relatively few phase-space configurations lead to product formation. In the collision complex "cauldron," particles form a near-ergodic distribution at short-range. The "cauldron" is very deep, supporting a huge number of possible configurations and implying a long, snarled trajectory through phase space.  In contrast, the exothermicity is small, leaving outgoing products with low kinetic energy and yielding only a small window of possible momenta with escape velocity.  Almost all of the energy must then be spent on sending the particle in the radial direction in order to form products.  Any energy spent on motion perpendicular to the radial direction threatens to rob the particle of escape velocity and to doom the particle to return to make another long, snarled pass through the cauldron.  This momentum window is a narrow "angle of acceptance" arising from the low exothermicity of the reaction (see Fig.~\ref{fig:Energetics}).  The momentum-space angle of acceptance and the position-space boundary of the short-range interaction region together define a phase-space bottleneck.  As the flux of products escaping through the bottleneck is much lower than the volume of phase-space in the collision complex, a long time is required for product formation from the collision complex.

Calculation of the collision complex decay rate in the reduced-dimensional system presented here is complicated by the type of interactions in the cauldron and the barrierless potential experienced by the products.  The method of "Gaussian bumps," previously applied in the field of quantum chaos,\cite{Heller,Bies.2001.016204,Luukko.2016.37656}  can be used to mimic the many-body interactions in the full-dimensional reaction, but leads to numerical problems, as the same exponential protrusions in the potential that induce chaos in the system also lead to numerical instability upon integration of the equations of motion. Degradation of energy conservation likewise occurs in alternative methods to induce chaos or randomness such as the kicked rotor,\cite{Casati.1979.334,Chirikov.1979.263,Izrailev.1980.417,Izrailev.1980.553} Fokker-Planck-Kolmogoroff equation,\cite{Fokker.1914.810,Planck.1917.324,Kolmogoroff.1931.415} and the Langevin equation.\cite{Langevin.1908.530}  Additionally, reactions are typically considered to proceed through a narrow range of position-space configurations in forming products.  Investigation of an activated complex or a position-space bottleneck is required to determine the reaction rate with traditional methods such as the Van't Hoff-Arrhenius law,\cite{VantHoff.1884.333,Arrhenius.1889.96,Arrhenius.1889.226}  the Lindemann-Hinshelwood mechanism,\cite{Lindemann.1922,Hinshelwood.1926} Rice-Ramsperger-Kassel-Marcus (RRKM) theory\cite{Rice.1927, Kassel.1928,Marcus.1951,Marcus.1952}, Transition State Theory (TST),\cite{Eyring.1935,Evans.1935,Wigner.1938,Miller.1974.1823,Laidler.1983} and phase-space theory.\cite{Light.1964.3221,Pechukas.1965.3281,Light.1965.3209,Nikitin.1965.90,Nikitin.1965.144,Chesnavich.1977.2306}   Even in theories that consider phase-space bottlenecks, saddle points are often required.\cite{Davis.1985,Davis.1986,Uzer.2002}  The phase-space bottleneck in the barrierless system presented here therefore necessitates a new method of analysis.  

This paper presents a new method of momentum kicks that addresses both of these concerns.  First, to remedy the computational difficulties involved in modeling the short-range interactions, a new method of energy-preserving momentum kicks is introduced in the two-dimensional system.\footnote{We thank Prof. Jack Wisdom of MIT for suggesting this and consulting with us on this work.}  The method of momentum kicks yields a near-ergodic distribution while ensuring energy conservation by employing "momentum kicks" in which the momentum vector is rotated by a random angle at regular time intervals in an inner deflection region within the potential.  In the two-dimensional system, the momentum kicks (momentum vector rotations) yield the same types of deflections expected in multi-atom collisions in the cauldron, and are expected to produce the same near-ergodic distributions. Second, to determine the collision complex decay rate for the barrierless system analytically, inspiration is taken from Wannier's method of phase-space counting.\cite{Wannier.1953.817}  As in Wannier's method, the first step is to locate the configurations that lead to product formation.  The ratio of the flux through the bottleneck to the phase-space volume of the cauldron then yields the analytical rate of collision complex decay.  

Poincaré surfaces of section provide a way to illustrate both the type of distribution arising from the momentum kicks and the existence of the phase-space bottleneck to product formation.  Birkhoff coordinates are ideal for illustrating the phase-space bottleneck.  Originally formulated for billiard problems,\cite{Birkhoff.1927} the momentum in Birkhoff coordinates relates the angle between a particle's position and momentum vectors, and is therefore directly proportional to the angle of acceptance.  Additionally, the numerical rate of collision complex decay is determined by monitoring the rate of product formation in numerical simulations of the two-dimensional system.  This rate can then be directly compared to the analytical rate.  We compare the analytical and numeric rates to verify the new classical method for collision complex decay in the reduced-dimensional model of the ultracold $\text{KRb}$ dimer reaction as a forerunner to its application full-dimensional ultracold systems.

\section{Methods}

\subsection{Post-threshold regime}\label{sec:PostThreshold}

To establish whether classical mechanics is applicable, we assess whether the two requirements of quantum-classical correspondence are met for collision complex decay in the two-dimensional model of the ultracold $\text{KRb}$ dimer reaction in which product dimers are taken to be structureless point particles.

First, we consider whether the semiclassical and classical probabilities agree.  Application of the stationary phase approximation to Feynman's path integral formulation of quantum mechanics yields the Van Vleck-Morette Gutzwiller propagator, which expresses the semiclassical probability as a sum over classical paths.\cite{VanVleck.1928.178,Morette.1951.848,Gutzwiller.1990,Heller} When there is only one classical path between initial and final states, the sum has only one term, and the semiclassical amplitude correctly reproduces the classical amplitude.  The Van Vleck-Morette-Gutzwiller propagator is
\begin{equation}
\left<x^\prime\middle|x(t)\right> \propto \sum_j\sqrt{\left|\frac{\partial^2S_j\left(x,x^\prime,t\right)}{\partial x \partial x^\prime}\right|}\exp\left(\text{i}S_j/\hbar\right),
\end{equation}
where $\hbar$ is the reduced Planck's constant, $S_j$ is the classical action of pathway $j$, and the term under the square root is related to the classical probability that a particle with sharp initial position $x$ evolves to overlap with a sharp final state at $x^\prime$ within time $t$, assuming no focal points are met along the way (\textit{i.e.}, where the accumulated phase and the Maslov index is zero $\nu_j=0$). For the long-range interaction between the products of the ultracold $\text{KRb}$ dimer reaction, the initial momentum and position uniquely define the path of a particle, as only one path satisfied the Euler-Lagrange equations connecting initial and final positions.  There is therefore only one classical path along the reaction coordinate connecting the collision complex to the products in the ultracold $\text{KRb}$ dimer reaction, such that the above semiclassical sum contains only a single term, the amplitude of the classical path.  The semiclassical amplitude then corresponds to the classical amplitude.

Second, we consider whether the semiclassical and quantum wavefunctions agree.  We term the regime above the threshold regime the "post-threshold" regime.\cite{Forrey.1999.2657,Mody.2001,Carraro.1995}  The post-threshold regime is defined by satisfaction of the criterion for the WKB approximation \cite{Wentzel.1926.518,Kramers.1926.828,Brillouin.1926.24,Brillouin.1926.353,Cote.1996,Mody.2001}
\begin{align}
1 & \gg \frac{1}{2\pi}\left|\frac{\text{d}\lambda(x)}{\text{d}x}\right|, \\
& \gg \left|\frac{\hbar}{p(x)}\frac{\text{d}p(x)}{\text{d}x}\right|, \\
& \gg \frac{m\hbar}{(p(x))^3}\left|\frac{\text{d}V(x)}{\text{d}x}\right|,\label{eq:WKBCriterion}
\end{align}
where $\lambda(x)=2\pi\hbar/p(x)$ is the de Broglie wavelength of the particle at position $x$ with momentum $p(x)$ in potential $V(x)$.  As the energy released in the ultracold $\text{KRb}$ dimer reaction is $14\text{ K}$ and the potential is given by the long-range interaction potential along the reaction coordinate described in Section \ref{sec:Hamiltonian}, the WKB condition is satisfied at any position along the potential.  The semiclassical wavefunction then closely approximates the quantum wavefunction.\cite{Forrey.1999.2657,Mody.2001,Carraro.1995}

As the collision complex decay satisfies both conditions, classical results are expected to agree with the quantum mechanical results.

\subsection{Model Hamiltonian}\label{sec:Hamiltonian}

Since the interacting product dimers in the ultracold $\text{KRb}$ reaction collision complex are treated as structureless point particles, the system is equivalent to a single particle in a two-dimensional central force potential. The two-dimensional $\text{KRb}$ dimer reaction was chosen as the means to validate the classical method as it retains the technical difficulties posed by the hot collision complex "cauldron" and the cold products of the ultracold $\text{KRb}$ dimer reaction.  The model does not account for quantization of the energy of the outgoing products that would be required for simulation of the full four-atom reaction. We chose to study the two-dimensional problem because it faces one of the main obstacles of the ultracold $\text{KRb}$ dimer reaction: the contrast between the depth of the collision complex "cauldron" and the shallowness of the low exothermicity.  The other goal of quantization of the transitions of the outgoing products remains.

The Hamiltonian is then
\begin{equation}
\mathcal{H}\left(r;p_{r},p_{\theta}\right)=\frac{p_{r}^{2}}{2m}+\frac{p_{\theta}^{2}}{2mr^{2}}+V(r),
\end{equation}
where $r$ is the radius, $p_{r}$ is the radial momentum, $p_{\theta}$ is the angular
momentum, $m$ is the reduced mass of the products, and $V(r)$ is the potential energy along the reaction coordinate $r$. 

The potential energy is given by the long-range interaction between the outgoing products. Since the products are considered to be structureless point particles and since the products are non-polar homonuclear molecules, the asymptotic interaction is isotropic and given by the inverse sixth-order van der Waals potential.\cite{Ospelkaus.2010,Marinescu.1994,Kotochigova.2006,Ni.2010,Kotochigova.2010}  The potential is parametrized to remove the singularity at the origin, match the potential well depth to the transition-state well depth, and ensure the potential reaches its asymptote at a physically realistic distance, which yields the long-range interaction potential
\begin{equation}
V(r)=-\frac{C_{6}}{\left(\beta r^{2}+\alpha\right)^{3}},\label{eq:Potential}
\end{equation}
where $C_{6}$ is the van der Waals dispersion coefficient and $\alpha$ and $\beta$ are parametrization constants.  The long-range potential is shown in Fig.~\ref{fig:Energetics}.  For the well minimum to correspond to the energetic favorability of the transition state at the bottom of the "cauldron", the origin $r=0$ is defined as the equilibrium transition-state distance between the forming products.  Particles confined to the inner regions of the central force potential are defined as non-products and particles able to pass to infinity $r\rightarrow\infty$ are defined as products.  The short-range attractive and repulsive multi-center interactions between the dimers are represented by the simplified model of energy-preserving momentum kicks.

\begin{figure}[H]
\begin{centering}
\includegraphics[width=0.50\textwidth]{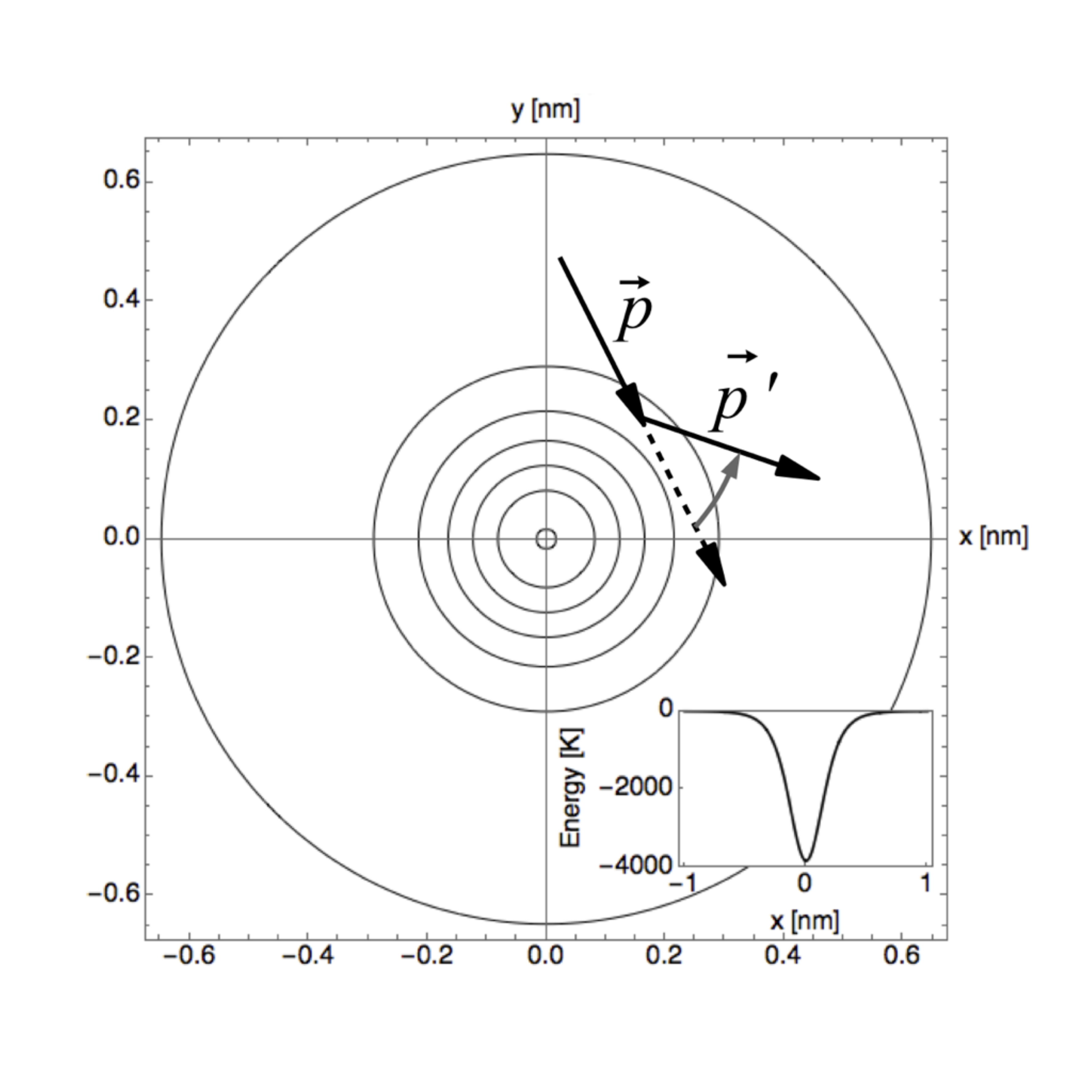}\includegraphics[width=0.50\textwidth]{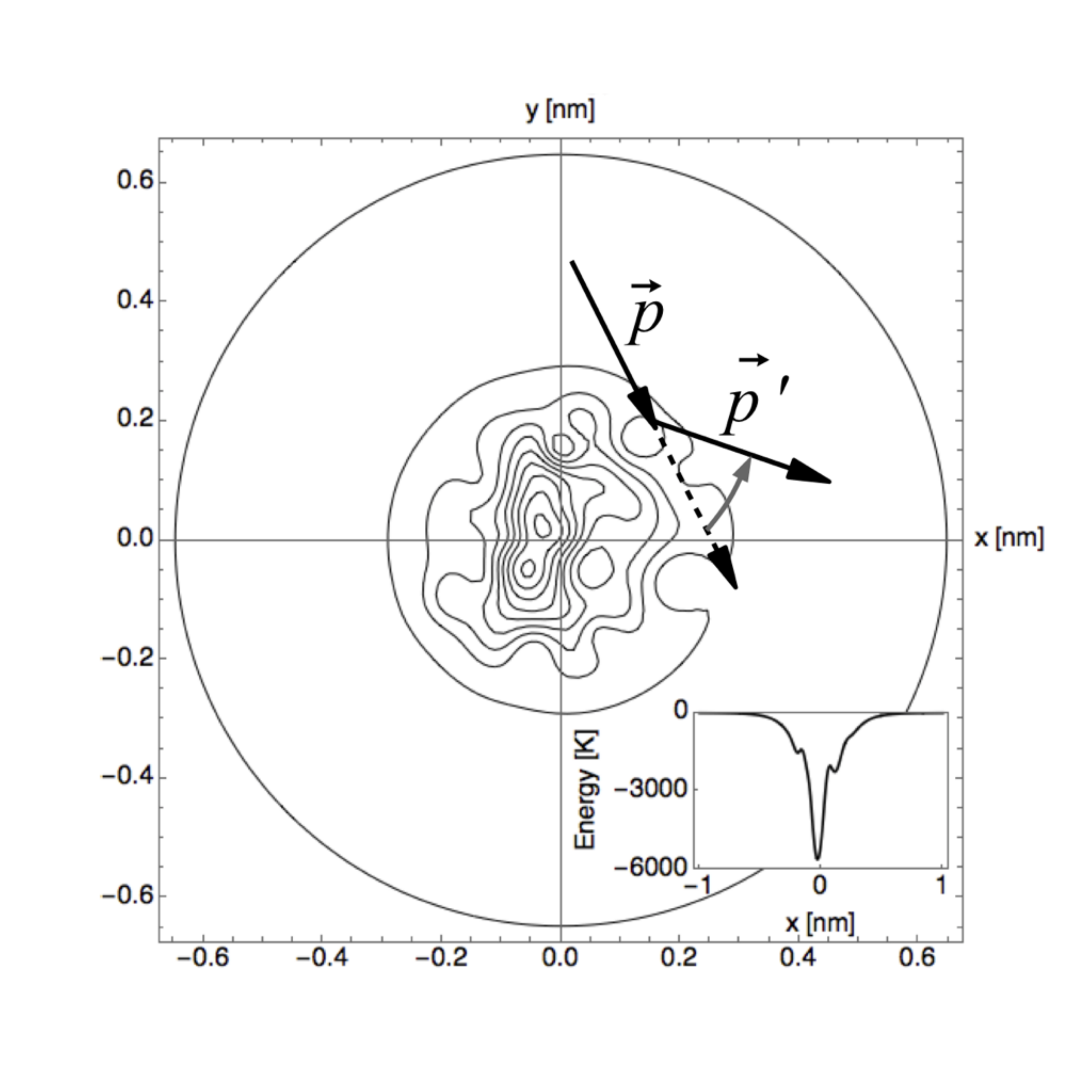}
\par\end{centering}
\caption{ (a) Momentum kicks in the method of energy-preserving momentum kicks deflect the momentum vector equivalently to (b) Gaussian bumps in the method of Gaussian bumps.\cite{Heller,Bies.2001.016204,Luukko.2016.37656} The momentum vector $\vec{p}$ (black solid and dashed arrows) is deflected (gray curved arrow) to the momentum vector $\vec{p}^\prime$ (black solid arrow). Linear contours of equal potential energy are shown.  Inset displays potential energy on $y=0$ slice. \label{fig:BumpsandKicks}}
\end{figure}

To employ the energy-preserving momentum kicks in the "cauldron", the momentum vector is rotated by a random angle at regular time intervals for particles within a distance $R$ along the reaction coordinate.  The values of the maximum angle of the kick, the time interval between kicks, and the limit of the deflection region boundary were chosen to model the response of the particle to protrusions in the potential Eq.~\ref{eq:Potential} arising from short-range interactions, as depicted in two dimensions in Fig.~\ref{fig:BumpsandKicks} with an analogy to the Gaussian protrusions of the Gaussian bump method.\cite{Heller,Bies.2001.016204,Luukko.2016.37656}

Since the energy-preserving momentum kicks rotate the momentum vector in the "cauldron" and the Hamiltonian has no explicit dependence on the angle, the angular momentum $p_\theta$ varies inside and is conserved outside the deflection region.  As the energy-preserving momentum kicks are energy-conserving and the Hamiltonian $\mathcal{H}$ is independent of the time $t$, the total energy $E$ is always conserved. As the energy of the incoming reactants is negligible compared to the exothermicity, the total energy of the system $E$ is approximately equal to the energy released in the reaction ($E=14\text{ K}$), the asymptotic value of the kinetic energy of the outgoing products.

\subsection{Determination of phase-space bottleneck\label{sec:Escape-criteria}}

\subsubsection{Critical momenta in polar coordinates\label{sec:ClassicalEscape}}

To determine the location of the bottleneck through which all outgoing products must pass, we determine which trajectories will reach infinite radius $r\rightarrow\infty$.  To find these trajectories, the critical radii and momenta are determined for passage over an extremum in the effective potential energy $V_\text{eff}$, where the effective potential $V_\text{eff}$ is composed of the sum of the centrifugal term and the potential energy.  The critical radii and momenta can then be determined in the same way that the maximum impact parameter for scattering is determined in classical capture theory,\cite{Langevin.1905.245,Gorin.1938.513,Gorin.1939.256} via location of the effective potential energy maximum.  Specifically, when the potential energy is given by Eq.~\ref{eq:Potential} and the parametrization parameter $\alpha$ is negligible, the critical energy $E_\text{crit}$ at which the effective potential is maximized is
\begin{align}
E_{\text{crit}} & =\frac{\beta^{3/2}\left|p_{\theta}\right|^{3}}{3^{3/2}2^{1/2}C_{6}^{1/2}m^{3/2}},
\end{align}
the critical radius $r_\text{crit}$ at which the critical energy $E_\text{crit}$ is reached is
\begin{align}
r_{\text{crit}} & =\sqrt[4]{\frac{6mC_{6}}{p_{\theta}^{2}\beta^{3}}}\label{eq:rcrit},
\end{align}
and the critical momenta $p_{r,\text{crit}}$ and $p_{\theta,\text{crit}}$ are the maximum angular momentum and the minimum radial momentum a particle can have and still pass through the critical radius $r_\text{crit}$
\begin{align}
p_{\theta,\text{crit}} & =3^{1/2}2^{1/6}C_{6}^{1/6}m^{1/2}E^{1/3}\beta^{-1/2}\label{eq:pthetacrit},\\
p_{r,\text{crit}} & =\sqrt{2mE+2mC_{6}\left(\beta r^{2}+\alpha\right)^{-3}-p_{\theta,\text{crit}}^2r^{-2}}.\label{eq:prcrit}
\end{align}
The turning point radius $R_\text{turn}$ then gives the maximum radius $r$ that can be reached by particles that cannot pass through the critical radius $r_\text{crit}$
\begin{equation}
R_{\text{turn}}=\sqrt[6]{\frac{2C_{6}}{E\beta^{3}}}.\label{eq:TurningRadius}
\end{equation}
The critical values are determined only for the potential of the outgoing products, as only the potential of the outgoing products is required to study the process of collision complex decay.

\subsubsection{Critical momentum in Birkhoff coordinates}\label{sec:Birk}

\begin{figure}[H]
\begin{centering}
\includegraphics[width=0.5\textwidth]{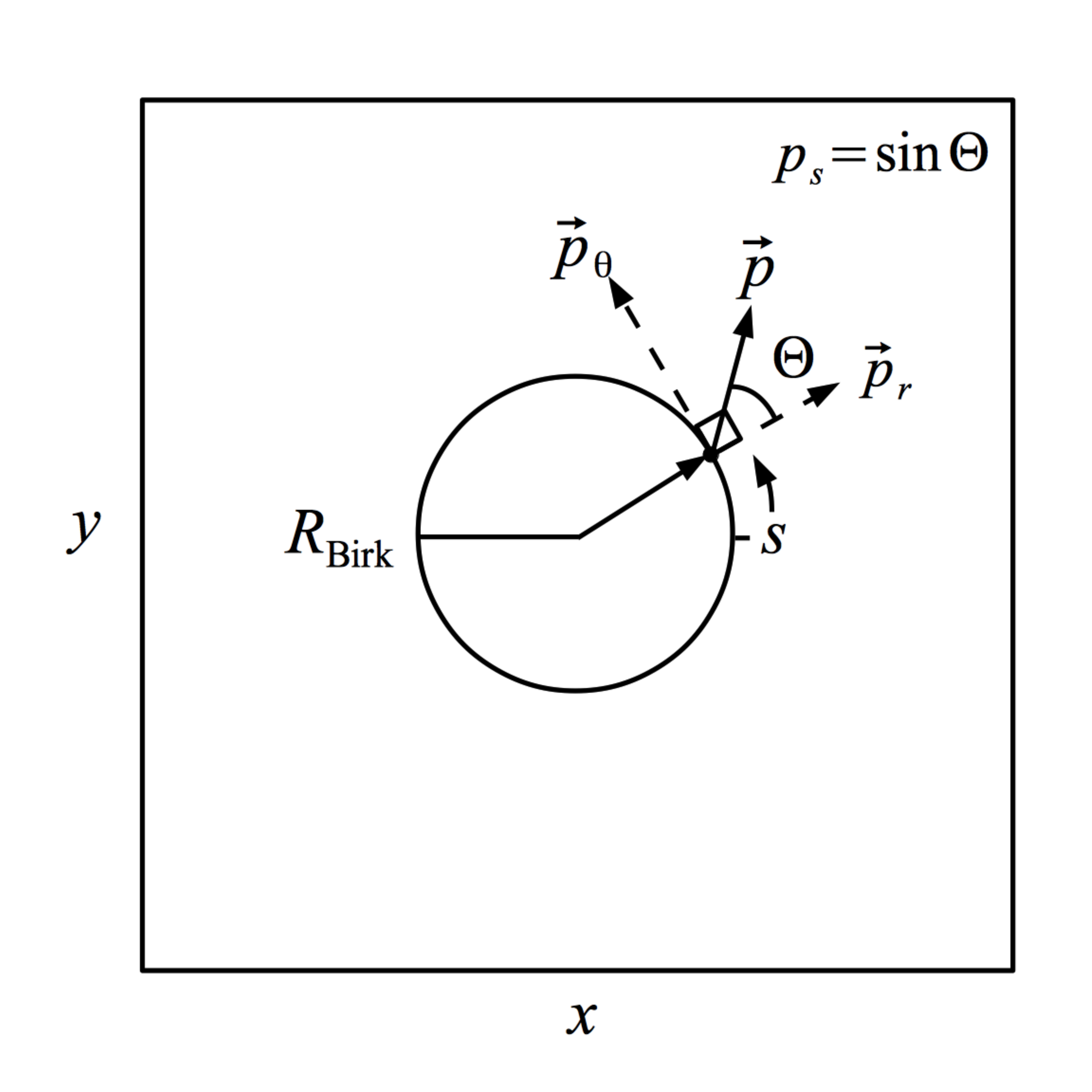}
\par\end{centering}
\caption{Birkhoff coordinates $(s,p_s)$ at Birkhoff radius $R_\text{Birk}$.\label{fig:Model-potential-energy-surface}}
\end{figure}

We can also specify the bottleneck in terms of Birkhoff coordinates.  To specify the phase-space bottleneck that separates non-products from products in Section \ref{sec:Bottleneck} in terms of the angle between a particle's position and momentum vectors, we employ Birkhoff coordinates  (see Fig.~\ref{fig:Model-potential-energy-surface}).\cite{Birkhoff.1927} An annulus of a fixed radius $r=R_{\text{Birk}}$ is considered about the origin of the central force potential.  The coordinates of a particle at Birkhoff radius $R_\text{Birk}$ are given by the angle $s$ of the particle along the annulus and the momentum $p_{s}$ with which the particle hits the annulus (the sine of the angle $\Theta$ between the particle's position and the momentum vectors),  as follows:
\begin{align}
s & =\theta,\\
p_{s} & =\sin\left(\Theta\right),\\
\Theta & =\arctan\left(\frac{p_{\theta}}{R_\text{Birk}p_{r}}\right).\label{eq:BirkhoffAngle}
\end{align}

Given the critical momenta in polar coordinates Eqs.~\ref{eq:pthetacrit}-\ref{eq:prcrit}, the critical momentum in Birkhoff coordinates is 
\begin{equation}
p_{s,\text{crit}}=\sin\left[\arctan\left(\frac{p_{\theta,\text{crit}}}{R_\text{Birk}p_{r,\text{crit}}}\right)\right].\label{eq:BirkhoffCrit}
\end{equation}

To create surfaces of section, the area-preservation of the application of Birkhoff coordinates to the billiard problem\cite{Birkhoff.1927} is exploited.  To ameliorate the concern that the particle in the central force potential eventually escapes the confines delimited by the Birkhoff position $s$ whereas billiards stay within the boundaries, we posit a reflecting wall at a radius beyond the radii under study $R_\text{refl}\gg \lim_{t\rightarrow\infty}r(t)$ to maintain the area preservation of the map.

\subsubsection{Phase-space bottleneck in Birkhoff coordinates}\label{sec:Bottleneck}

To locate the phase-space bottleneck to collision complex decay, the surface through which product-forming trajectories that will pass unimpeded to infinite radius $r\rightarrow\infty$ is found.  Since the radius $r=R$ is the maximum radius at which the energy-preserving momentum kicks operate, such that no product particles are found within the radius $R$ and all product particles are found outside the radius $R$, the radius $R$ is the coordinate-space component of the bottleneck. Since the critical momentum $p_{s,\text{crit}}$ Eq.~\ref{eq:BirkhoffCrit} divides product particles that will pass to infinity $r\rightarrow\infty$ from those non-products trapped within the turning point radius $r\le R_\text{turn}$, the critical momentum $p_{s,\text{crit}}$ constitutes the momentum-space component of the bottleneck.  The complete phase-space bottleneck is then 
\begin{align}
r & =R,\\
|p_{s}| & < p_{s,\text{crit}}.
\end{align}
All product-forming particles must then pass through this bottleneck, with product-forming particles restricted to leave the deflection region at $r=R$ with a momentum $p_s$ within a narrow angle of acceptance. The angle of acceptance $\tilde\Theta$ spans over all Birkhoff angles $\Theta$ Eq.~\ref{eq:BirkhoffAngle} of product-forming particles. The critical Birkhoff angle $\Theta_\text{crit}=\arctan\left(\frac{p_{\theta,\text{crit}}}{R_\text{Birk}p_{r,\text{crit}}}\right)$ is the maximum Birkhoff angle that product-forming particles can have, such that the angle of acceptance is twice the critical Birkhoff angle $\tilde\Theta=2\Theta_\text{crit}$.

\subsection{Analytic rate}\label{sec:Rate}

To calculate the collision complex lifetime, the rate of product formation from the collision complex was determined.  To calculate the rate, we assume an ergodic population in which trajectories populate phase-space equally in the deflection region of the collision complex "cauldron". The assumption of ergodicity allows the rate constant $k$ to be calculated as a phase space average over the
statistical distribution of particles.\cite{vonNeumann.1932.263,Birkhoff.1931.656,Birkhoff.1942.222,Parry.1981,Petersen.1996.171}

\begin{figure}[H]
\begin{centering}
\includegraphics[height=0.3\textheight]{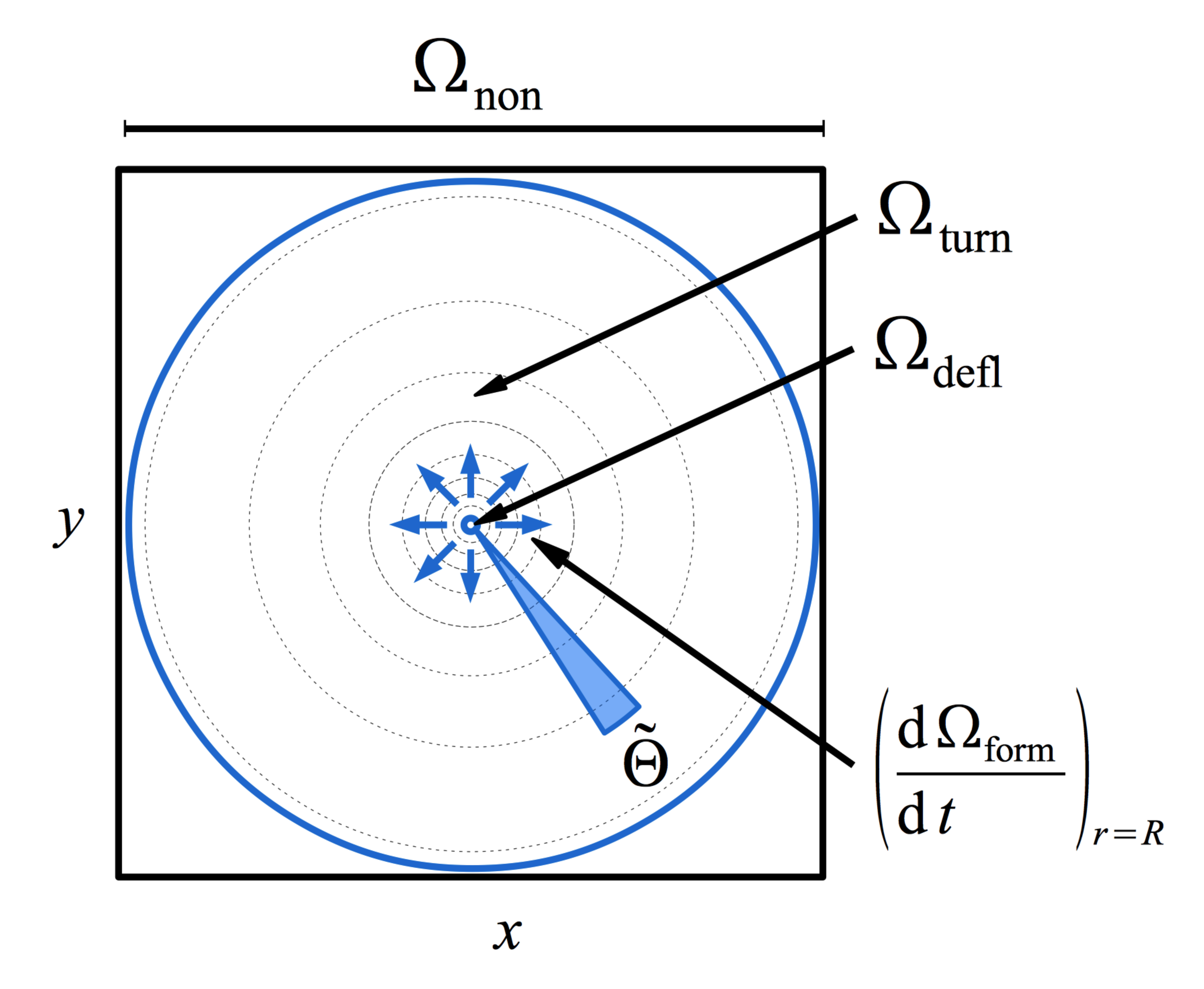}
\par\end{centering}
\caption{Schematic of calculation of the rate constant $k$ from the phase-space ratio Eq.~\ref{eq:ErgodicRateConstant} of the volume flux $\left(\frac{\text{d}\Omega_{\text{form}}}{\text{d}t}\right)_{r=R}$ to the volume of non-products $\Omega_\text{non}$ (the sum of the volumes of the deflection region $\Omega_\text{defl}$ and the turning region $\Omega_\text{turn}$). Products are formed when the trajectory crosses the deflection region boundary at radius $r=R$ within the angle of acceptance $\tilde{\Theta}$, shown magnified $20$ times for visibility.\label{fig:Calculations}}
\end{figure}

The ergodic rate $k$ is given by the ratio of the volume
flux $\left(\frac{\text{d}\Omega_{\text{form}}}{\text{d}t}\right)_{r=R}$ (the
phase-space volume $\Omega_{\text{form}}$ that the product-forming trajectories pass
through at radius $r=R$ in an infinitesimal time interval
$\text{d}t$) to the total phase-space volume of the non-products $\Omega_{\text{non}}$ (the
sum of the inner deflection $\Omega_{\text{defl}}$ and outer turning
region $\Omega_{\text{turn}}$ volumes)
\begin{equation}
k=\frac{\left(\frac{\text{d}\Omega_{\text{form}}}{\text{d}t}\right)_{r=R}}{\Omega_{\text{non}}}.\label{eq:ErgodicRateConstant}
\end{equation}
A schematic of the phase-space ratio is shown in Fig.~\ref{fig:Calculations}.

\subsubsection{Volume flux}

To determine the flux of forming products passing through the phase-space bottleneck, we employed the phase-space counting method of Wannier.\cite{Wannier.1953.817}. The phase space volume is given by integration over the coordinates
\begin{equation}
\Omega=\int\int\int\int\text{d}r\text{d}\theta\text{d}p_{r}\text{d}p_{\theta}.\label{eq:PhaseSpaceVolume}
\end{equation}
Given conservation of the total energy $E$, the volume flux is reformulated in terms of integration over the radial momentum $p_{r}$ and divided by an infintesimal time increment $\text{d}t$ to yield the volume flux through the phase-space bottleneck in Section \ref{sec:Bottleneck}
\begin{align}
\left(\frac{\text{d}\Omega_{\text{esc}}}{\text{d}t}\right)_{r=R} & =\int_{0}^{2\pi}\text{d}\theta\int_{-p_{\theta,\text{crit}}}^{p_{\theta,\text{crit}}}\text{d}p_{\theta},\\
 & =4\pi{p_{\theta,\text{crit}}},\label{eq:VolumeFlux}
\end{align}
Given the energy dependence of the critical angular momentum $p_{\theta,\text{crit}}$ Eq.~\ref{eq:pthetacrit}, the energy-dependence post-threshold law of the volume flux is $E^{1/3}$.

\subsubsection{Rate constant}

To calculate the overall rate of collision complex decay, the total phase-space volume of non-products is given by the total volume within the phase space bottleneck in Section \ref{sec:Bottleneck}, the sum of the phase-space volumes of
the inner deflection region $\Omega_{\text{defl}}$ in which the particles are
randomly deflected and the outer turning region $\Omega_{\text{turn}}$ in
which the particles leave the deflection region only to reenter.  

To calculate the total phase-space volume of non-products, the inner region is considered to be delimited by the maximum deflection radius $r=R$. The outer region is considered to reside between the boundary of the deflection region $R$ and the maximum turning point radius $R_\text{turn}$ in Eq.~\ref{eq:TurningRadius}, and contains all momenta outside of the phase-space bottleneck in Section \ref{sec:Bottleneck}.  The total phase-space volume Eq.~\ref{eq:PhaseSpaceVolume} of non-products is then
\begin{align}
\Omega_{\text{non}} & =m\int_{0}^{R}\text{d}r\int_{0}^{2\pi}\text{d}\theta\left(2\int_{0}^{p_{\theta,\text{max}}}\frac{\text{d}p_{\theta}}{p_{r}}\right)+m\int_{R}^{R_\text{turn}}\text{d}r\int_{0}^{2\pi}\text{d}\theta\left(2\int_{p_{\theta,\text{crit}}}^{p_{\theta,\text{max}}}\frac{\text{d}p_{\theta}}{p_{r}}\right)\\
 & =m\pi^{2}R^{2}+\pi^{2}m\left(R_\text{turn}^2-R^{2}\right)-4\pi m\left(F\left(R,R_\text{turn}\right)\right).\label{eq:Total}
\end{align}
where $F(a,b)$ is the integral
\begin{equation}
F\left(a,b\right)=\int_a^b r\arctan\left(\frac{p_{\theta,\text{crit}}}{rp_{r,\text{crit}}}\right)\text{d}r.\label{eq:FIntegral}
\end{equation}

The volume flux Eq.~\ref{eq:VolumeFlux} and phase-space volume
Eqs.~\ref{eq:Total} together yield the ergodic
rate constant Eq.~\ref{eq:ErgodicRateConstant}
\begin{equation}
k=\frac{4p_{\theta,\text{crit}}}{m\pi{R_\text{turn}^2}-4mF\left(R,R_\text{turn}\right)}.\label{eq:AnalyticRateConstant}
\end{equation}
As the system is in the post-threshold regime described in Section \ref{sec:PostThreshold}, the energy-dependence constitutes the post-threshold rate law, in contrast to the threshold rate law that would be observed in the quantum regime.  At high energies at which the turning radius $R_\text{turn}$ Eq.~\ref{eq:TurningRadius} is approximately equal to the boundary of the inner deflection region $\Omega_\text{turn}$, the non-analytic integral $F\left(R,R_\text{turn}\right)$ is negligible.  Expression of the critical angular momentum $p_{\theta,\text{crit}}$ Eq.~\ref{eq:pthetacrit} and the turning radius $R_\text{turn}$ Eq.~\ref{eq:TurningRadius} in terms of the energy $E$ then reveals an energy dependence of $E^{2/3}$.

\subsection{Numerical rate}

To corroborate the results of the new classical method, the analytic rate was compared to the numerical rate.  To calculate the numerical rate, a classical simulation of collision complex break-up was performed for the two-dimensional model of the ultracold $\text{KRb}$ dimer reaction described in Section \ref{sec:Hamiltonian}.  To model the separation of the outgoing products, the mass $m$ of the particle in the simulation was given by the reduced mass of the product $\text{K}_{2}$ and $\text{Rb}_{2}$ dimers.  To simulate the system being initialized in the collision complex, particles were initialized at a small radius, radius $r_{\text{init}}\le1\text{ a.u.}$.  The particle trajectory was then integrated with Velocity Verlet with a time step $\tau=0.01\text{ a.u.}$ (Hartree atomic units $m_{e}=e=\hbar=k_{\text{e}}=1\text{ a.u.}$) chosen to be sufficiently small to ensure energy conservation within a fraction of a percent.  The particles' coordinates in phase-space were then recorded at short time intervals to produce smooth images of the particle trajectories.  The coordinates were recorded every $2^{12}$ time steps near the deflection region ($r\le1.5R$) in which the deep potential well led to relatively high velocities.  The coordinates were then recorded every $2^{16}$ time steps in the asymptotic regions where velocities were lower.  Particles were considered as having formed products once they passed through the phase-space bottleneck described in Section \ref{sec:Bottleneck}.  To ensure the full process of collision complex break-up was simulated, the trajectory of each particle was simulated until it reached well past the maximum radius reachable by non-product particles $R_{\text{turn}}$ Eq.~\ref{eq:TurningRadius}.  The turning point radius was $R_{\text{turn}}=18\text{ a.u.}$ at the exothermicity of the reaction $E=14\text{ K}$.

To model the pseudo-one-dimensional process of product separation described in Section \ref{sec:Hamiltonian}, the potential energy of the particle was given by the parametrized inverse sixth-order van der Waals potential Eq.~\ref{eq:Potential}.  The parameters were chosen to reflect both the characteristic depth of the cauldron and the  characteristic distance of the van der Waals interaction to produce a realistic depiction of the reaction energetics along the reaction coordinate $r$.  Without the availability of the van der Waals dispersion coefficient for the products, the van der Waals dispersion coefficient for the $\text{KRb}-\text{KRb}$ reaction $C_{6}=16130\text{ a.u.}$\cite{Marinescu.1994,Kotochigova.2006} was employed to give a constant of the appropriate order. The characteristic length parameter $\beta$ was chosen to yield a weak potential once the outgoing dimers were at a significant distance from each other, as the van der Waals interaction becomes weak where the particles are well separated. The value of the parameter $\beta=2.9\text{ a.u.}$ was chosen to yield a full width at half maximum (FWHM) equal to the predicted equilibrium $\text{K-Rb}$ distance at the CCSD(T)\cite{Purvis.1982.1910} level of theory.\cite{Byrd.2010} To ensure the well depth was equal to the predicted transition-state well depth at the CCSD(T) level of theory, the normalization constant was $\alpha=110.\text{ a.u.}$.  This choice of the well depth ensures that the relatively high kinetic energies and the phase-space volume in the potential well reflect those possible in the collision complex cauldron.  To ensure the WKB criterion was satisfied at the energy $E=14\text{ K}$, the quantity in Eq.~\ref{eq:WKBCriterion} was determined to be less than one at all positions, reaching a maximum at $0.48$ at $11\text{ a.u.}$. To determine the difficulty of product formation, the angle of acceptance at the energy $E=14\text{ K}$ was found to be $\tilde\Theta=1.2\text{ rad}$.

At short-range, random energy-preserving momentum kicks were employed to efficiently mimic the many-body interactions in the cauldron.  As these interactions are strongest where the atoms are nearby in the collision complex cauldron, the maximum deflection radius $R$ was placed at a region where the interactions between separating dimers were expected to be sufficiently weak.  For the purpose of the simulation, the maximum deflection radius was chosen to be at a distance where the intermolecular distance had stretched to $1.5$ times the equilibrium bond distance, $R=4.\text{ a.u.}$.  To model the gradual effects of the interactions, the kicks were chosen such that particles in the deflection were smoothly deflected as if by weak perturbations in the potential.  To ensure the deflection was smooth, the angle of the momentum rotation was chosen to be small ($\theta_\text{defl}=\left[-\frac{\pi}{10},\frac{\pi}{10}\right]\text{ rad}$) and the kicks were chosen to occur at the same time interval at which dynamics were recorded (every $2^{12}$ time steps). To encourage formation of the near-ergodic ensemble in the deflection region that would arise naturally from four-body interactions, the strength of the kicks was chosen to be independent of the position of the deflection region.  The robustness of the method was verified by measuring the complex lifetime for various kick angles, kick intervals, and kick strength with radii.

For comparison of the numerical and analytic post-threshold rate laws, simulations were carried out for microcanonical ensembles at a range of energies spanning several orders of magnitude.  In addition to simulation at the predicted product energy of the ultracold $\text{KRb}$ dimer reaction, simulations were also carried out at other temperatures to verify the analytical energy-dependence of the rate.  Low-energy simulations were included despite being in the quantum reflection threshold regime for the sole purpose of establishment of the classical energy-dependence of the rate.  For total system energies of $E=0.01\text{ K}$, $E=0.1\text{ K}$, and $E>0.1\text{ K}$; $2000$, $2048$, and $8192$ particles were simulated; respectively. At higher energies, more particles were simulated to yield a significant population of remaining non-products after the premature formation of products by transient particles was excluded. Results were only analyzed for times after which $1024$ or fewer particles remained to ensure ergodicity had been reached.  Exponential regression of the number of non-product particles remaining over time yielded the numerical ergodic rate constant $k$ for collision complex break-up.  For direct comparison, the numerical rate constant was plotted against the analytical rate constant, which was computed as a power regression for sampled energies in the range $E\in[10^{-7},800]\text{ K}$.

\section{Results}

\begin{figure}[H]
\begin{centering}
\includegraphics[height=0.3\textheight]{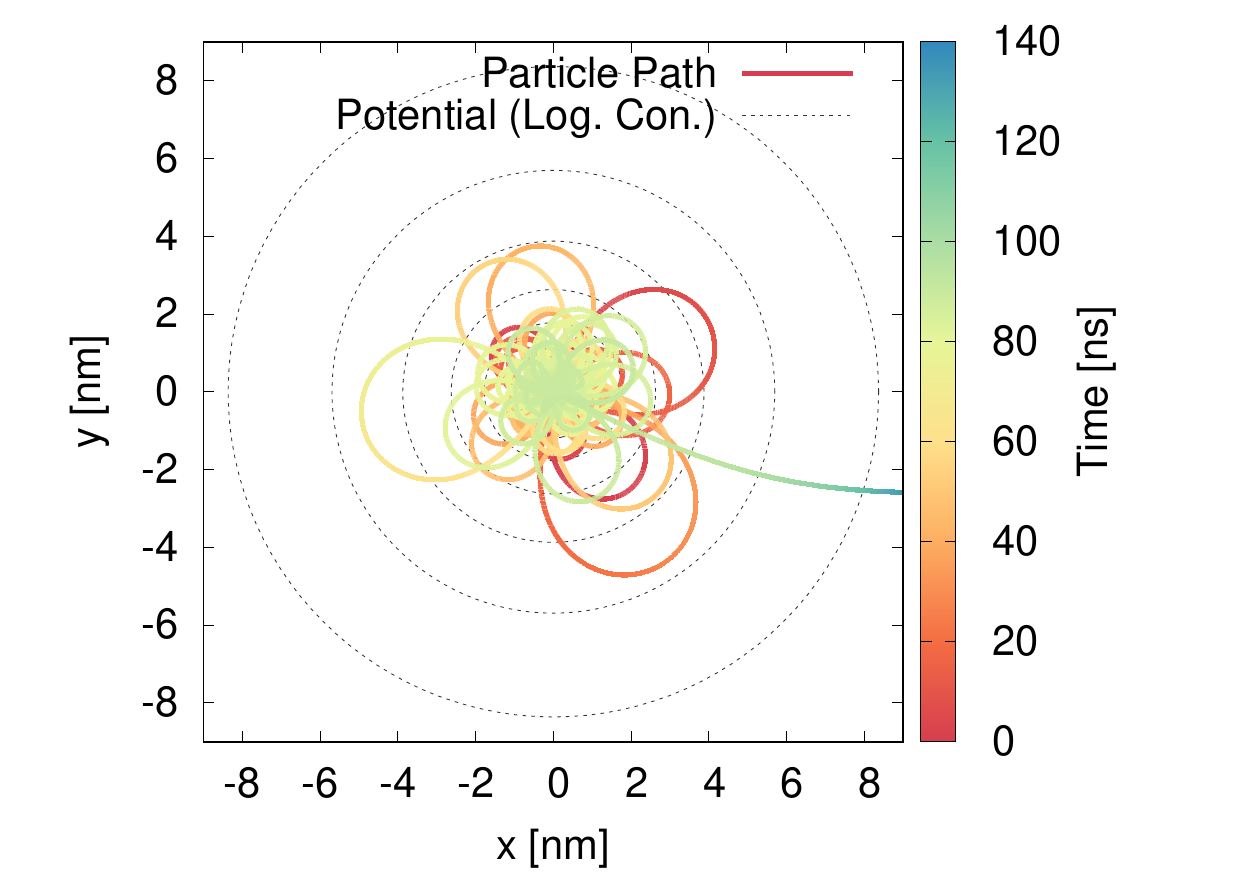}
\par\end{centering}
\caption{Sample particle trajectory (solid multicolor line) in potential Eq.~\ref{eq:Potential} (dashed gray logarithmic contours) simulated numerically.\label{fig:Trajectory}}
\end{figure}

The particles in the numerical simulation exhibited snarled trajectories associated with a collision complex.  A representative particle path is shown in Fig.~\ref{fig:Trajectory}.  The trajectory is shown at a low energy $E=10^{-5}\text{ K}$ to illustrate a difficult "escape."  In all of the individual trajectories, the particle moved randomly within the deflection region, and, upon reaching the deflection region boundary, made an attempt to leave the deflection region.  Two behaviors were evidenced in these attempts.  In failed escape attempts ("non-product") to leave the deflection region, the particle returned to the deflection region.  In successful escape attempts ("product") the particle left the deflection region without returning.  In the trajectory shown in Fig.~\ref{fig:Trajectory}, the particle repeatedly made failed escape attempts until the particle left the deflection region with escape velocity.  Successful escape attempts passed through the phase-space bottleneck described in Section \ref{sec:Bottleneck} while failed escape attempts did not.  The behavior matched the expected behavior of the collision complex, in which the collision complex is expected to sample the many possible configurations in the collision complex "cauldron," making failed attempts to form products when the energy is outside of the phase-space bottleneck, until it reaches a configuration in the phase-space bottleneck that allows it to irreversibly form products.

\begin{figure}[H]
\begin{centering}
\includegraphics[width=1.\textwidth]{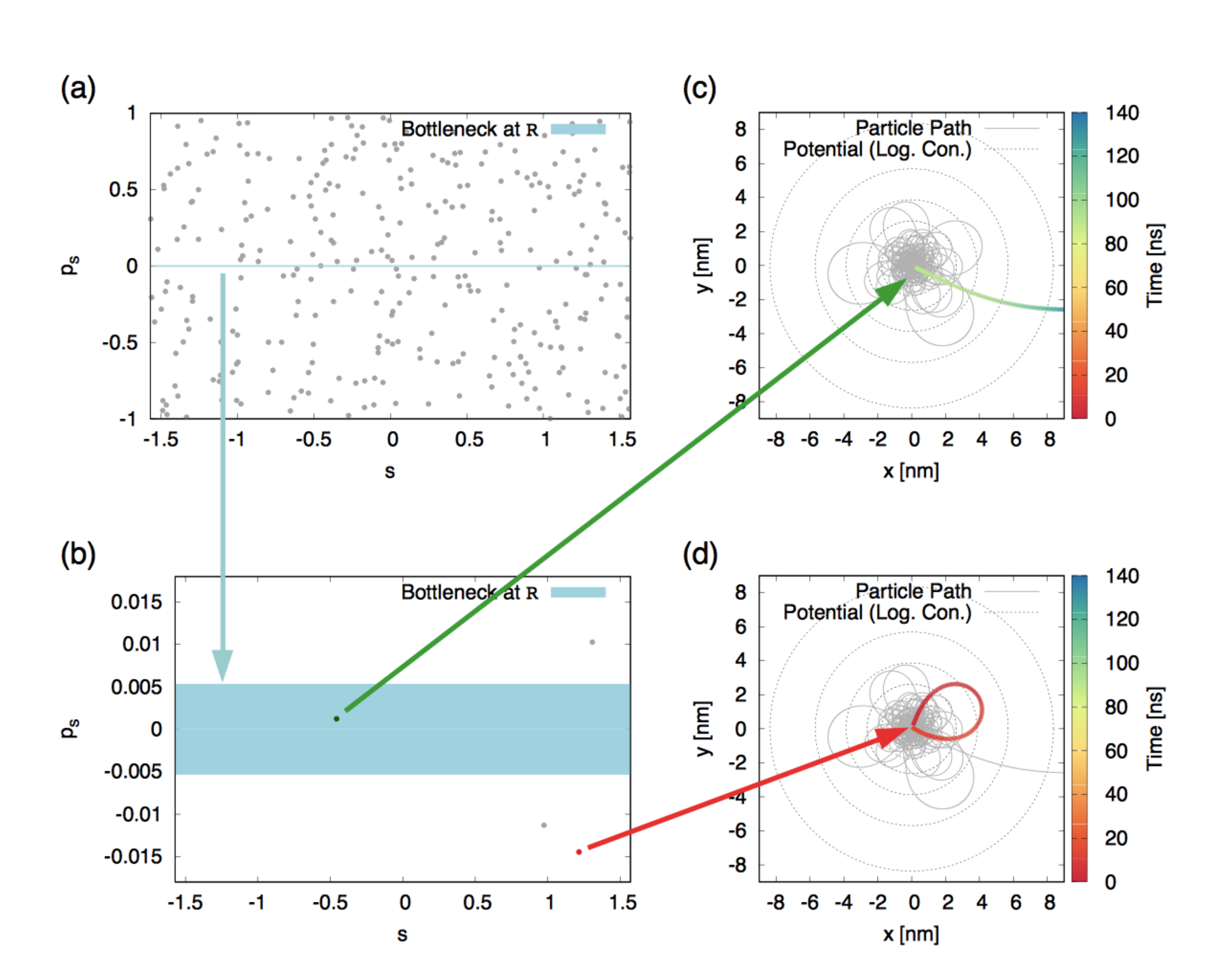}
\par\end{centering}
\caption{(a) Evenly-spaced points (light gray) on the Poincaré surface of section for the particle shown in Fig.~\ref{fig:Trajectory} ($R_\text{Birk}=4.\text{ a.u.}$, $p_{r}>0$) reflect the near-ergodicity of the system. (b) Two points are highlighted as examples of a successful product escape (green) and a failed non-product escape (red) in a x56 magnified image of the phase-space bottleneck (blue rectangle).  (c-d) Trajectory plots illustrate the portions of the trajectory (thick multicolor line) that contributed the highlighted points on the surface of section.
\label{fig:Surface-of-section-sample-particle}}
\end{figure}

The phase-space bottleneck was also reflected in surfaces of section taken at the boundary of the deflection region.  A representative Poincaré surface of section is shown in Fig.~\ref{fig:Surface-of-section-sample-particle} for the trajectory in  Fig.~\ref{fig:Trajectory}.  The points were spaced evenly throughout the surface of section with only one passage through the phase-space bottleneck.  The uniformity of the point spacing in Fig.~\ref{fig:Surface-of-section-sample-particle}(a) was in keeping with that expected for a near-ergodic system. The even spacing reflected the random behavior arising from the energy-preserving momentum kicks. The surface of section did not exhibit saddle points, in contrast to applications of phase-space transition state theory.\cite{Wigner.1938,Davis.1985,Davis.1986}  In addition, the single passage through the bottleneck was in keeping with irreversible formation of products.

Passage through the Birkhoff radius $R_\text{Birk}$ was associated with successful and failed escape attempts on the trajectory, dependent on whether or not the Birkhoff momentum $p_s$ fell inside the bottleneck described in Section \ref{sec:Bottleneck}.  When the particle had more than the critical Birkhoff momentum $p_{s,\text{crit}}$ Eq.~\ref{eq:BirkhoffCrit}, the particle made a failed attempt to escape the potential, and the particle left the deflection region only to return.  When the particle had less than the critical Birkhoff momentum $p_{s,\text{crit}}$ Eq.~\ref{eq:BirkhoffCrit}, the particle passed through the phase-space bottleneck to make a successful escape attempt, and the particle left the deflection region and ultimately passed to the maximum radius included in the simulation, well past the turning radius $R_\text{turn}$ Eq.~\ref{eq:TurningRadius} that bounded non-product trajectories at the energy $E$ of the simulation.  The behavior agreed with what would be expected for a system with a phase-space bottleneck separating the collision complex "cauldron" from outgoing products.

\begin{figure}[H]
\centering{}\includegraphics[width=0.5\textwidth]{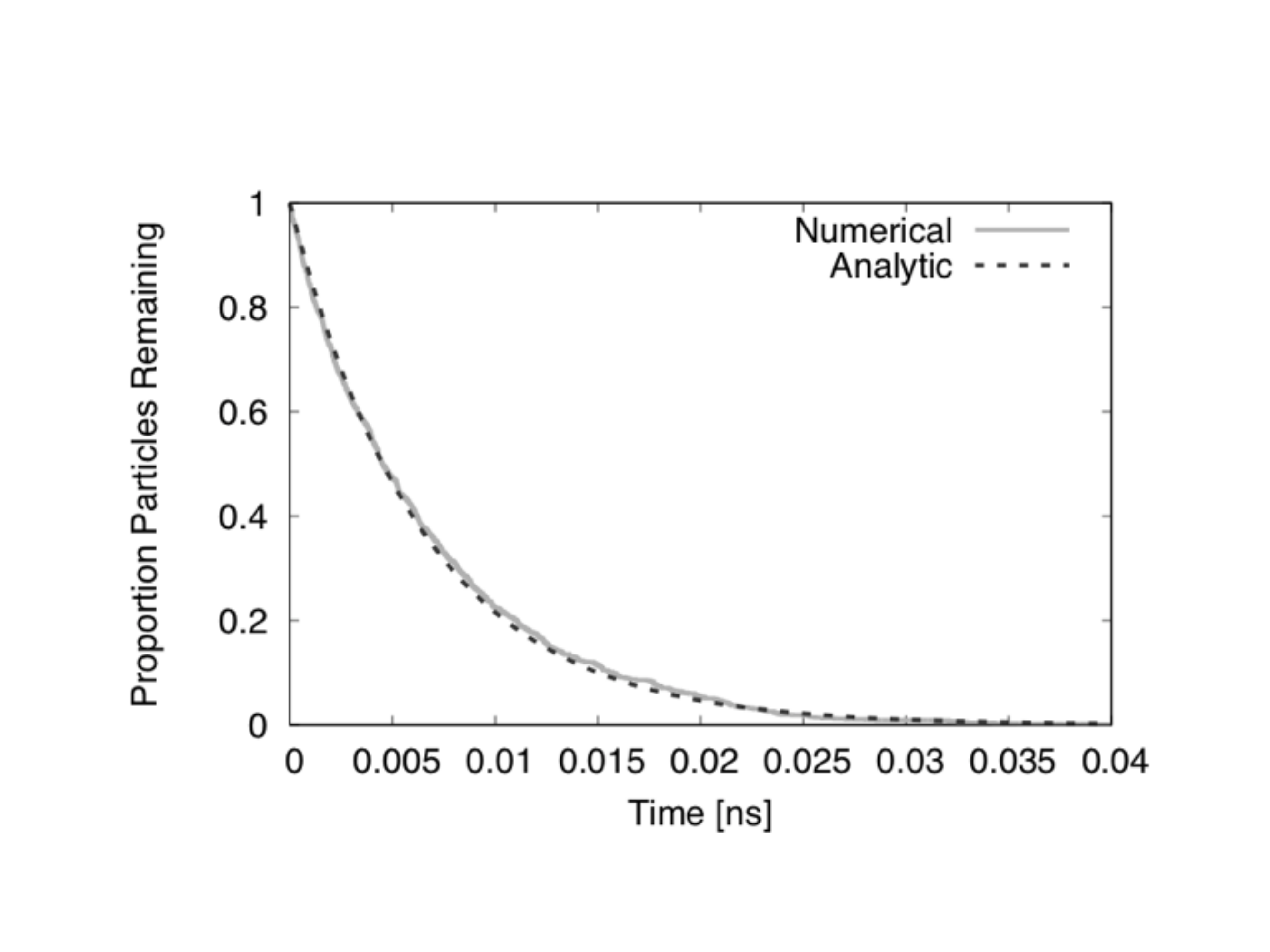}\centering{}\includegraphics[width=0.5\textwidth]{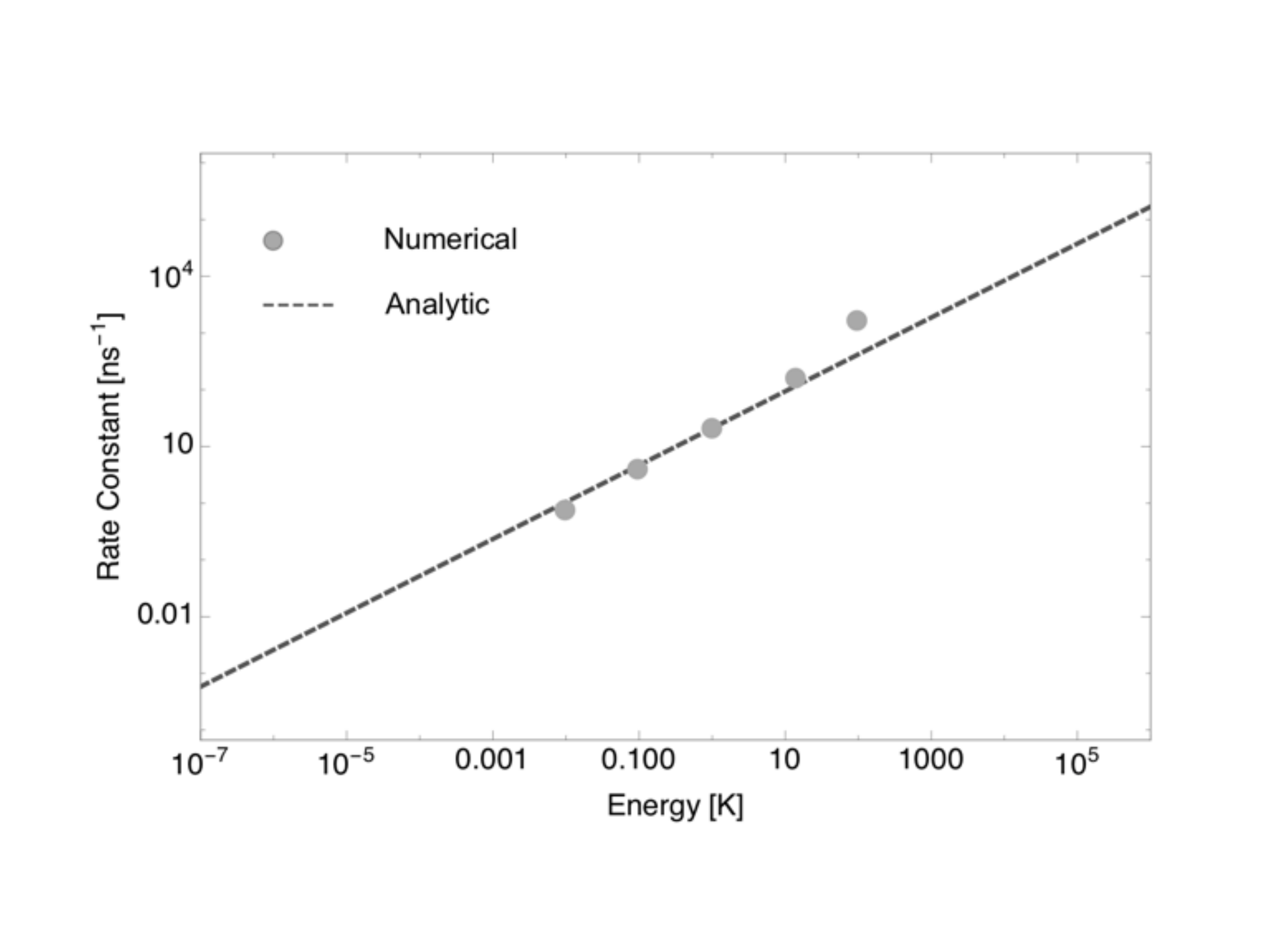}
\caption{(a) Remaining proportion of non-products over time at $14\text{ K}$ calculated numerically (solid light grey line) and analytically (dashed dark grey line). (b) Rate constant $k$ of product formation determined numerically (light grey points) and analytically (dashed dark grey lines).  Maximum standard asymptotic error was $0.03\%$, indistinguishable from the point at the plot scale.\label{fig:Remaining-proportion-of-particles}}
\end{figure}

Observation of the proportion of non-product particles remaining in the collision complex cauldron over time in the numerical simulation closely agreed with the proportion predicted to remain analytically.  The close agreement at the energy released in the reaction, $E=14\text{ K}$, is shown in Fig.~\ref{fig:Remaining-proportion-of-particles}(a). The numerical rate constants were found to match the analytical rate constants at various energies over several orders of magnitude, as shown in Fig.~\ref{fig:Remaining-proportion-of-particles}(b).

Discrepancies in the calculated rate constants may be attributed to limitations in the methods of calculation of the analytic and numerical rate constants. Accurate analytic calculation of the rate constant $k$ was limited by numerical errors in evaluation of the non-analytic integral Eq.~\ref{eq:FIntegral} and the assumption that the normalization constant $\alpha$ is negligible in Section \ref{sec:ClassicalEscape}.  Accurate numerical calculation of the rate constant $k$ was also limited by variations in the relaxation time in which near-ergodicity is achieved.

The agreement between the analytic and numerical values for the rate constant $k$ was also evident in the half-life $t_{1/2}$, the characteristic timescale for disappearance of half of the remaining non-products.  As the half-life $t_{1/2}$, inversely related to the rate constant $k$, provides a measure of the collision complex lifetime, agreement between the values suggested the value of the methods to study collision complex lifetimes in sufficiently exothermic ultracold reactions.

At the energy released in the ultracold \text{KRb} reaction $E=14\text{ K}$, the analytic rate constant was $k=153\text{ ns}^{-1}$ (half-life $t_{1/2}=4.5\text{ ps}$) and the numerical rate constant was $k=149.$ (half-life $t_{1/2}=4.6\text{ ps}$, asymptotic standard error $\triangle k=\pm0.025$).  The timescale was sub-millisecond as in the Rice-Ramsperger-Kassel-Marcus (RRKM) transition state theory\cite{Rice.1927,Kassel.1928,Marcus.1951,Marcus.1952} treatment of the transition state decay in the full $\text{KRb}-\text{KRb}$ reaction, although the the timescales differ significantly and both methods are qualitative in their current form.\cite{Mayle.2013.012709} The RRKM analysis predicted a RRKM lifetime of the collision complex of $\tau=3.5\text{ }\upmu\text{s}$ for the total angular momentum $J=0$ and $J=1$ states, $\tau=3.6\text{ }\upmu\text{s}$ for the $J=2$ state, and $\tau=3.7\text{ }\upmu\text{s}$ for the $J=3$ state. The results of the RRKM method and the present method are not directly comparable as the RRKM method does not include the effects of a phase-space bottleneck.  In addition, the present method does not include consideration of the total angular momentum $J$ or the internal structure of the dimers.  The application presented here is only a partial solution of the reaction, as the products are considered to be structureless point particles.

\section{Discussion}

Our semiclassical argument for quantum-classical correspondence suggests that classical mechanics provides a new window into the study of collision complex decay in sufficiently exothermic ultracold reactions. Whereas quantum mechanical methods become intractable for ultracold systems of many heavy atoms, causing difficulties even in the two-dimensional case, the classical methods presented here help bring the study of complex ultracold chemical reactions within reach.  The successes of our method for a two-dimensional model of collision complex decay in the ultracold $\text{KRb}$ dimer reaction suggest that classical mechanics can be used as an essentially exact alternative to quantum mechanics for specific elements of collision complex decay in sufficiently exothermic ultracold reactions.

Our introduction of a simplified model of many-body interactions takes advantage of this classical picture to simulate the decay process efficiently.  Whereas other methods can lead to numerical instability, application of classical energy-preserving momentum kicks in the two-dimensional system is computationally economical and satisfies energy conservation inherently.  Agreement between the analytic ergodic rate law, determined with Wannier's method of phase-space counting,\cite{Wannier.1953.817} and the numerical rate law, determined with energy-preserving momentum kicks, supports the validity of this new method for inducing a near-ergodic distribution and determining the rate in a barrierless reaction.  The success of the method of energy-preserving momentum kicks and Wannier phase-space counting demonstrated here in a reduced-dimensional simulation of the ultracold $\text{KRb}$ dimer reaction, in which products were treated as structureless point particles, bodes well for generalization to full-dimensional simulations.

Our fully classical method not only enables computational simulations of collision complex decay in sufficiently exothermic ultracold reactions, but also provide intuition as to how these processes proceed.  Instead of visualizing the interaction of wavefunctions, the reaction can be visualized in terms of classical particles that must proceed through a narrow window in momentum space in order to form products.  In this picture, since the exothermicity of the reaction is low, the particle must focus almost all its available momentum on direct radial separation in order to form products, as without escape velocity the particle would reenter and remix in the collision complex "cauldron."  

Replacement of point particles with structured particles will require quantized vibrations and rotations and subsequently the addition of quantum effects.  We acknowledge that classical tools as presented here do not yet provide essentially exact results for the full-dimensional reaction. Instead, the classical techniques presented here give a classical view into collision complex decay in sufficiently exothermic ultracold chemical reactions and a means to calculate aspects of the process nearly exactly where quantum mechanics presents difficulties.  The study successfully illustrates the idea that collision complex decay in sufficiently exothermic ultracold reactions can be approached from a classical perspective.

\section*{Acknowledgements}

M. S. acknowledges financial support from the National Science Foundation Graduate Research Fellowship under Grant No. DGE1144152 and the Harvard Graduate School of Arts and Sciences Merit/Graduate Society Term-time Research Fellowship. We would like to thank Professor K.-K. Ni for discussions on ultracold chemistry, Professor J. Wisdom for insight concerning dynamical systems, Professor A. Aspuru-Guzik for stimulating discussions,  Professor V. S. Batista for support, and an anonymous reviewer for invaluable suggestions.

\bibliographystyle{unsrt}
\bibliography{KRb}

\end{document}